\documentclass[%
 aip,
 amsmath,amssymb,
 reprint,%
]{revtex4-1}

\usepackage{graphicx}
\usepackage{dcolumn}
\usepackage{bm}

\usepackage[utf8]{inputenc}
\usepackage[T1]{fontenc}
\usepackage{mathptmx}
\usepackage{siunitx}
\usepackage{gensymb}

\def\twod/{2D(3$v$)}
\def\um/{\textmu m} 
\newcommand{\Wcm}[2][0]{\ifnumcomp{0}{=}{#1}{}{$#1~\times$~}$10^{#2}$~W~cm$^{-2}$} 
\def\mildI/{\Wcm[5.4]{17}}
\def\relI/{\Wcm[3]{18}}

\begin{document}

\preprint{AIP/123-QED}

\title{Particle-in-Cell modeling of a potential demonstration experiment for double pulse enhanced target normal sheath acceleration}

\author{Nashad Rahman}
\email{rahman.176@osu.edu}
 \affiliation{Department of Physics, The Ohio State University, Columbus, OH, 43210, USA}
\author{Joseph R. Smith}
 \affiliation{Department of Physics, The Ohio State University, Columbus, OH, 43210, USA}
\affiliation{Department of Materials Science and Engineering, The Ohio State University, Columbus, OH, 43210, USA}
\affiliation{Air Force Institute of Technology, Dayton, OH, 45433, USA}
\author{Gregory Ngirmang}
\affiliation{National Academies of Sciences, Engineering, and Medicine, Washington DC, 20418, USA}
\affiliation{Air Force Research Laboratory, Dayton, OH, 45433, USA}

\author{Chris Orban}
 \affiliation{Department of Physics, The Ohio State University, Columbus, OH, 43210, USA} 

\date{\today}

\begin{abstract}
Ultra intense lasers are a promising source of energetic ions for various applications. An interesting approach described in Ferri et al. 2019 argues from Particle-in-Cell simulations that using two laser pulses of half energy (half intensity) arriving with close to 45 degrees angle of incidence is significantly more effective at accelerating ions than one pulse at full energy (full intensity). For a variety of reasons, at the time of this writing there has not yet been a true experimental confirmation of this enhancement. In this paper we perform 2D Particle-in-Cell simulations to examine if a milliJoule class, $5\cdot 10^{18}$ \si{W.cm^{-2}} peak intensity laser system could be used for such a demonstration experiment. Laser systems in this class can operate at a kHz rate which should be helpful for addressing some of the challenges of performing this experiment. Despite investigating a 3.5 times lower intensity than Ferri et al. 2019 did, we find that the double pulse approach enhances the peak proton energy and the energy conversion to protons by a factor of about three compared to a single laser pulse with the same total laser energy. We also comment on the nature of the enhancement and describe simulations that examine how the enhancement may depend on the spatial or temporal alignment of the two pulses.
\end{abstract}

\maketitle

\section{Introduction}
\label{sec:intro}

Sources of energetic ions can be useful for a variety of industrial, biomedical, and defense applications\cite{Bulanov2002_proton_therapy, Salamin_medical_ions,NVion,Tian_etal2018,ene2009applications,jain2011ion,pRad}. However, the size and cost of conventional accelerators can limit the use of this technology. Ultra-intense laser systems can potentially serve as a compact source of energetic ions. As discussed, for example, in \citet{Palmer2018} laser technology is maturing rapidly towards this goal.

There are still a number of technical challenges involved in translating proof-of-concept laser experiments to real-world applications. Among these challenges, the need to maximize the number of energetic ions accelerated by the laser interactions and to increase the peak energy of these ions is paramount. \citet{Ferri2019,Ferri_etal2020}  outline an approach to increase the peak ion energy and numbers of energetic ions (without increasing the total laser energy) by dividing the laser energy into two laser pulses that irradiate the target from different angles. The authors refer to this approach as ``enhanced Target Normal Sheath Acceleration" (enhanced TNSA). \citet{Ferri2019,Ferri_etal2020} find best results for these two pulses arriving at 45 degree angles of incidence with pulses arriving at 40 degrees and 50 degrees performing almost as well. Recently, \citet{Jiang_etal2020} provided simulation results describing a similar approach to enhanced TNSA using three laser pulses instead of two. In the present work, we only discuss double pulse enhanced TNSA for simplicity and because a three pulse experiment is even more difficult to perform than a two pulse experiment.

From an experimental point of view, there are a number of challenges with implementing the double pulse setup. In order for constructive interference to occur, both laser pulses need to arrive on target within the duration of the laser pulse. This can potentially be done using a delay line but it adds another element to the experiment that was not necessary for a single pulse. Once adequately synchronized, both micron scale laser spots need to spatially align despite instrumental vibrations. Perhaps because of these challenges, in their literature review \citet{Ferri2019} could only cite experimental papers where two pulses arrive on target from the same angle but with a time delay \cite{Markey_etal2010,Brenner_etal2014,Ferri_etal2018} and one paper in which an intense pulse reflects from the target and is redirected to the target area by reflecting from a spherical shell \cite{Scott_etal2012}. We note that \cite{Wang_etal2018} uses two pulses from different angles on two different targets in a multi stage approach to TNSA. 
However none of these experimental studies consider simultaneous double pulse irradiation of targets from two different angles of incidence like \cite{Ferri2019,Ferri_etal2020} propose.
The closest experimental work that we could find is \citet{Morace2019} who use the interference of four overlapping, picosecond timescale ``beamlets" to create a periodic interference pattern that shapes the target during the interaction to enhance ion acceleration. These beamlets are aimed at the target with a relatively large spot size and  small angles of incidence (a few degrees). That study finds enhanced ion acceleration but not nearly as much as \cite{Ferri2019,Ferri_etal2020} suggest is possible. Also, the mechanism that the authors propose is distinct from what \cite{Ferri2019} and \cite{Ferri_etal2020} describe and requires a longer timescale.

An appealing platform for demonstrating double pulse enhanced TNSA that we consider in this paper would be a milliJoule class short pulse laser system operating at a high repetition rate. Such an experiment would face a number of challenges. Instrumental vibrations will produce shot-to-shot variations in the position of the laser spots that could easily be larger than the width of the laser spot itself. An additional challenge would be ensuring that the laser pulses arrive within the duration of the pulse. But, if coupled to a data acquisition system capable of single shot analysis and a precision delay line, one could measure the ejected ion energy spectrum in rare events where both micron scale laser spots are well aligned spatially and temporally. Over time one can collect statistics that explore how the ejected proton spectrum depends on the distance between the spots and temporal alignment.

A particularly good example of a milliJoule class laser system with advanced diagnostics is the laser system used in \citet{morrison_etal2018}. As described in \citet{Morrison_etal2015}, 5~mJ laser pulses with a peak intensity near $5 \cdot 10^{18}$~W~cm$^{-2}$ were used to accelerate protons up to 2.5~MeV in energy and at a kHz repetition rate. As discussed in a perspectives article by \citet{Palmer2018}, producing $\sim$2~MeV protons from a few mJ class laser system operating at a kHz repetition rate had not been achieved before.

In the present work we perform Particle-in-Cell (PIC) simulations in an effort to determine if the laser system described by \citet{morrison_etal2018} could be a useful platform for demonstrating double pulse enhanced laser ion acceleration as described in \citet{Ferri2019,Ferri_etal2020}. A key question is whether the $5 \cdot 10^{18}$~W~cm$^{-2}$ peak intensity laser pulses produced by this system, if split into two pulses and directed to the target, will produce enhanced ion acceleration even though the intensities involved are 3.5 times lower than the lowest intensity described in \citet{Ferri2019} and a different target is used. Given that the nature of the effect is not yet perfectly understood (as will be discussed later), and that there can be intensity thresholds in this range for various other kinds of laser phenomena (e.g. \cite{Orban_etal2015,Smith_etal2019}), we can answer these questions with simulations rather than simply assuming that the double pulse approach will remain effective in the regime accessible through experiments with milliJoule class lasers. To provide additional guidance to our experimental colleagues, we also present additional simulation results that address how robust the enhanced proton energies are to spatial or temporal misalignment or phase offsets.

In Sec.~\ref{sec:setup} we describe our simulation setup. In Sec.~\ref{sec:results} we describe our simulation results for perfect temporal and spatial alignment of the two pulses and compare to the single pulse case. In Sec.~\ref{sec:scan} we present another set of simulation results that consider the impact of spatially or temporally misaligned pulses on the proton energies. In Sec.~\ref{sec:discussion} we discuss the simulation results in both Sec.\ref{sec:results} and Sec.~\ref{sec:scan} and comment on the nature of the effect. In Sec.~\ref{sec:conclusion} we conclude.

\section{Simulation Setup}
\label{sec:setup}

We performed 2D3v PIC simulations using the LSP code \cite{Welch_etal2004} to simulate intense laser interactions with a 0.46~\si{\micro\metre} thick Ethylene Glycol target. This target was irradiated with a either a single beam with intensity of $5\cdot 10^{18}$~\si{W.cm^{-2}} or two beams of intensity $2.5 \cdot 10^{18}$~\si{W.cm^{-2}}. All beams were 780~nm in wavelength and incident on the target p-polarized at 45\degree\  to normal and with a spot size of 2.2~\si{\micro\metre} (FWHM) and a pulse width of 42~fs Full Width at Half Maximum (FWHM) with a sine squared envelope. The single beam simulation (including laser and target) was overall similar to an experiment and simulation described in \citet{morrison_etal2018} which demonstrated proton energies up to 2.5~MeV.

We used the implicit field advance and particle advance in the LSP code with 2D3v dimensionality. Unless otherwise noted, the simulation grid was 2800 by 2800 cells with a physical size of 28~\si{\micro\metre} $\times$ 28~\si{\micro\metre}. We also performed two simulations that will be described that were twice this physical size with twice as many cells in each direction in an effort to determine if the finite size of the target was biasing our results.

Each cell that contained target material used 9 particles each of protons, electrons, singly ionized Oxygen, and singly ionized Carbon resulting in 36 total particles per cell. The target electron density was $10^{23}$ cm$^{-3}$ with other species in proportion to ensure that the target is neutral. Field ionization using the ADK model \cite{ADK1986} was included in the simulations to allow Carbon and Oxygen ions to be further ionized by either the laser electric field or the electrostatic fields on the target\cite{PPT1966,ADK1986}. No pre-plasma was assumed in the simulations we present in the body of this paper. Instead, there was only the maximum density or vacuum. In Appendix~\ref{ap:preplasma} we present results from simulations that do assume an exponential scale length preplasma. 
Throughout the initial temperature of the particles was set at 1~eV. These simulations were run for 1~ps with 20~as timesteps.

In Sec.~\ref{sec:results} we present simulations where the two laser pulses are perfectly aligned in terms of the laser spot and perfectly aligned temporally, meaning no time delay between the two pulses and the peak intensity reaches the target at the same time. In Sec.~\ref{sec:scan} we present results from a number of simulations that consider misaligned pulses as described in that section. In both sections we compare to single pulse simulations with full intensity and energy and the same temporal duration.

\begin{figure*}
\includegraphics[width=17cm]{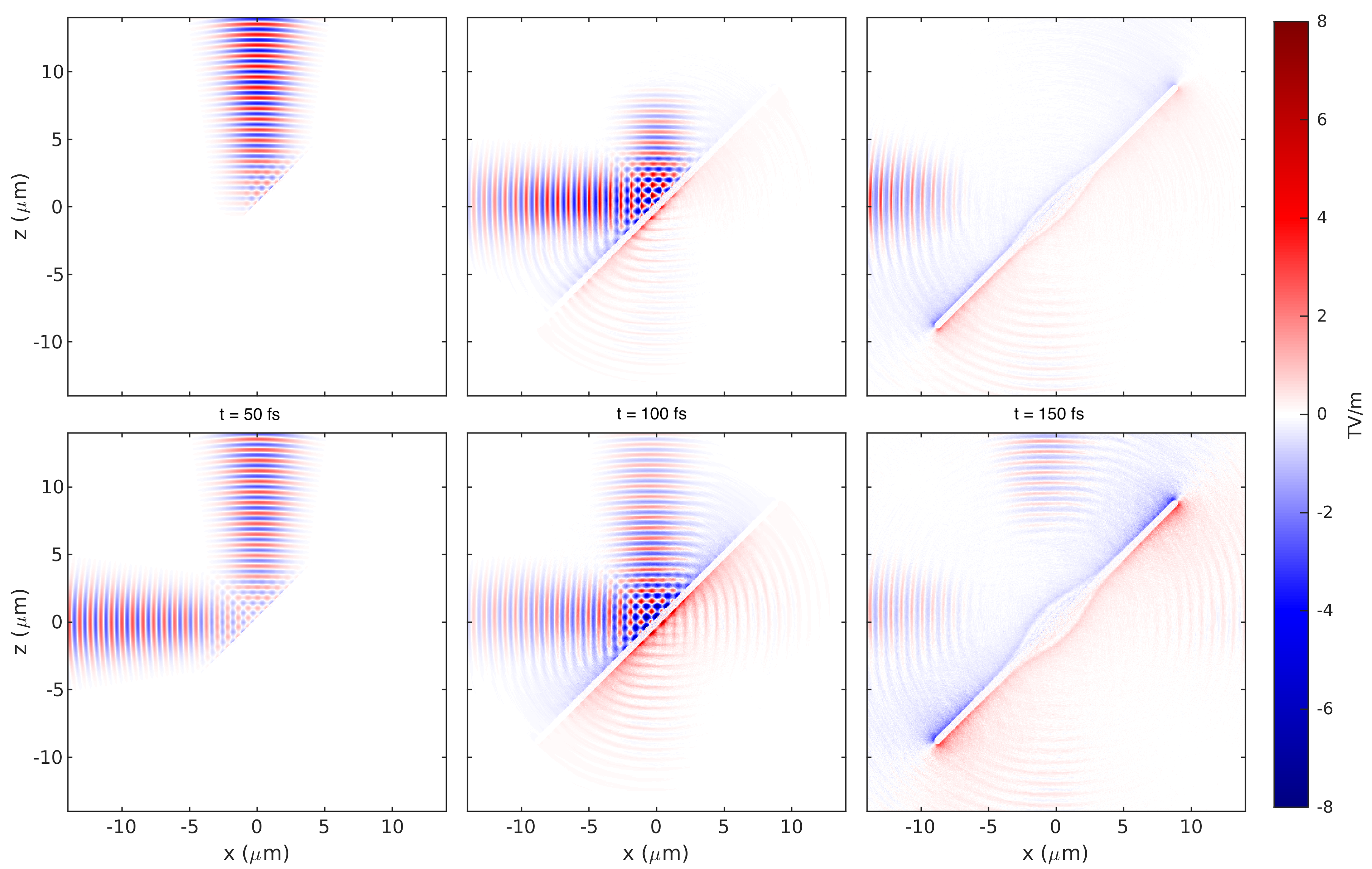}
\vspace{-0.5cm}
\caption{Electric field component normal to the target for the single pulse simulation (top panels) and double pulse simulation (bottom panels) at 50~fs (left panels), 100~fs (center panels), and 150~fs (right panels) }\label{fig:efields}
\end{figure*}

\begin{figure*}
\includegraphics[width=2.7in]{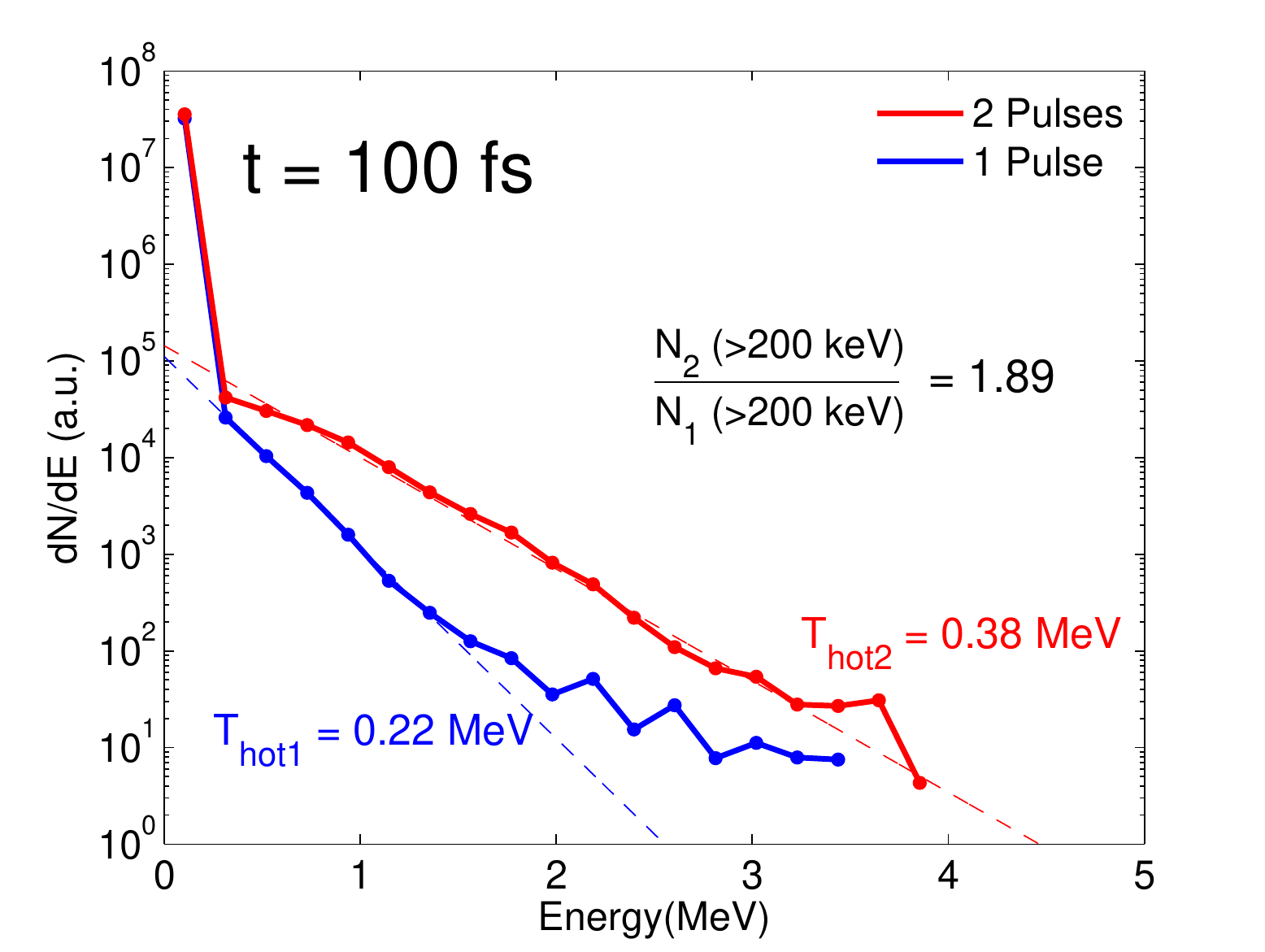}\includegraphics[width=2.7in]{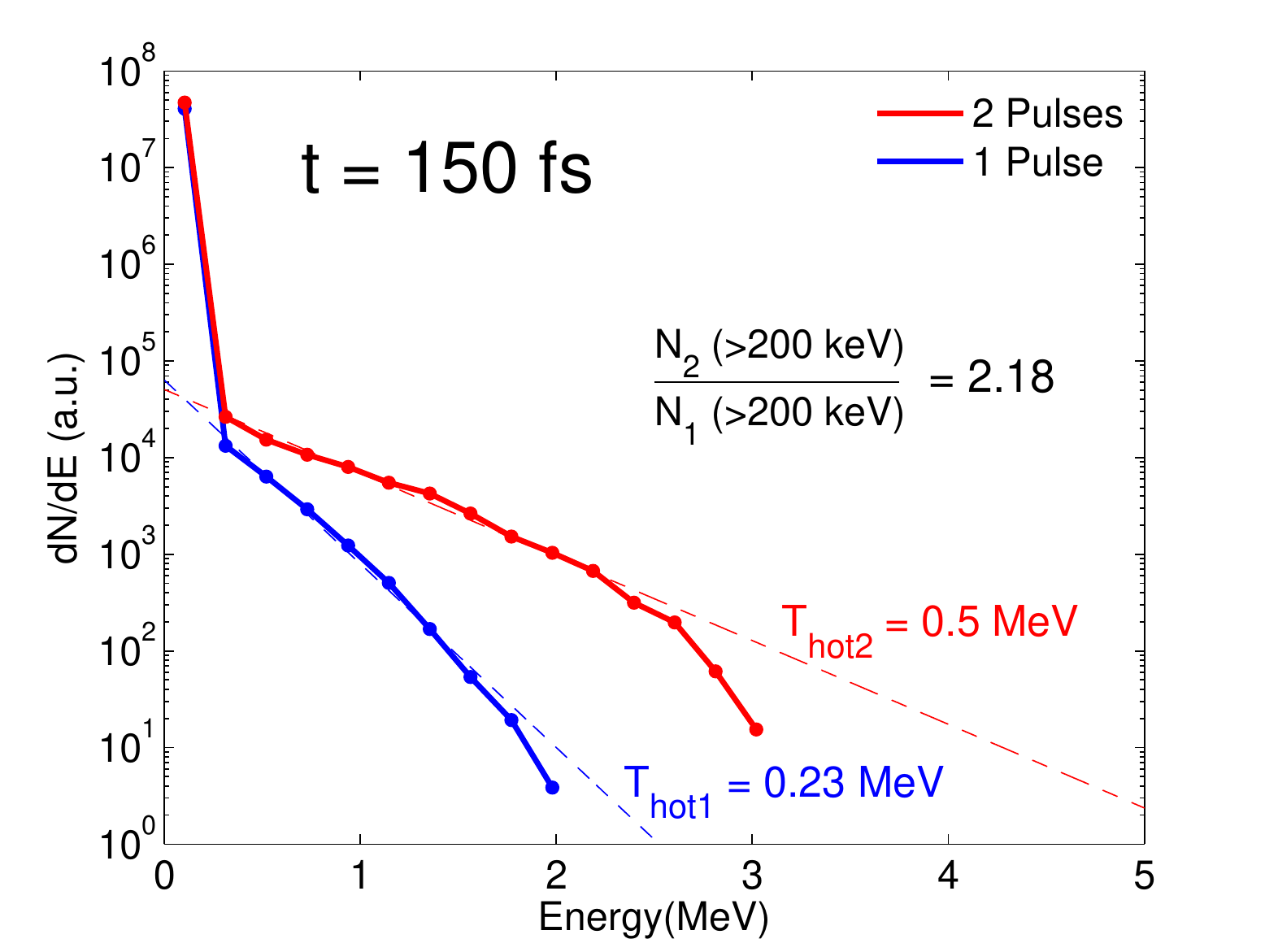}
\includegraphics[width=2.7in]{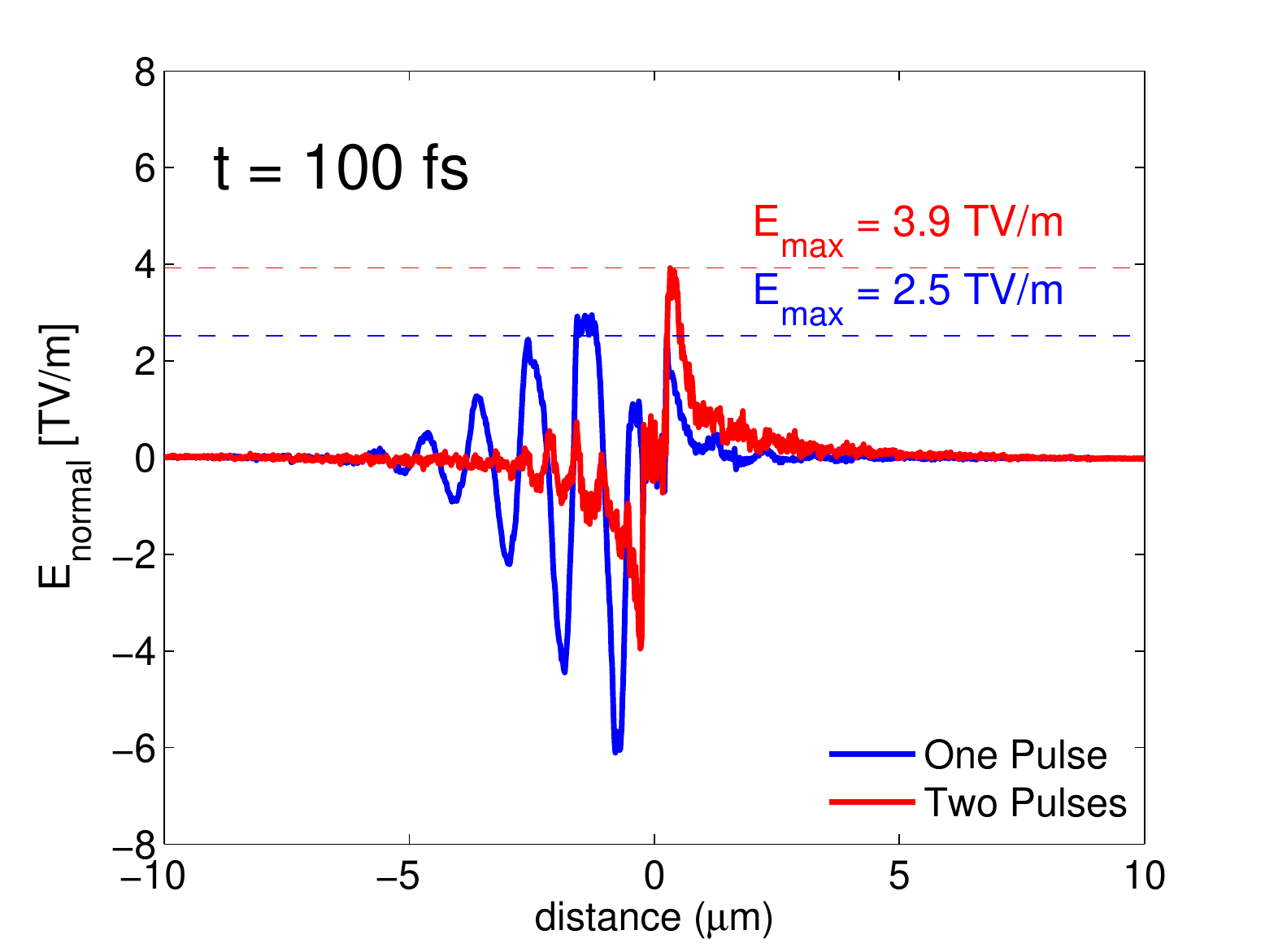}\includegraphics[width=2.7in]{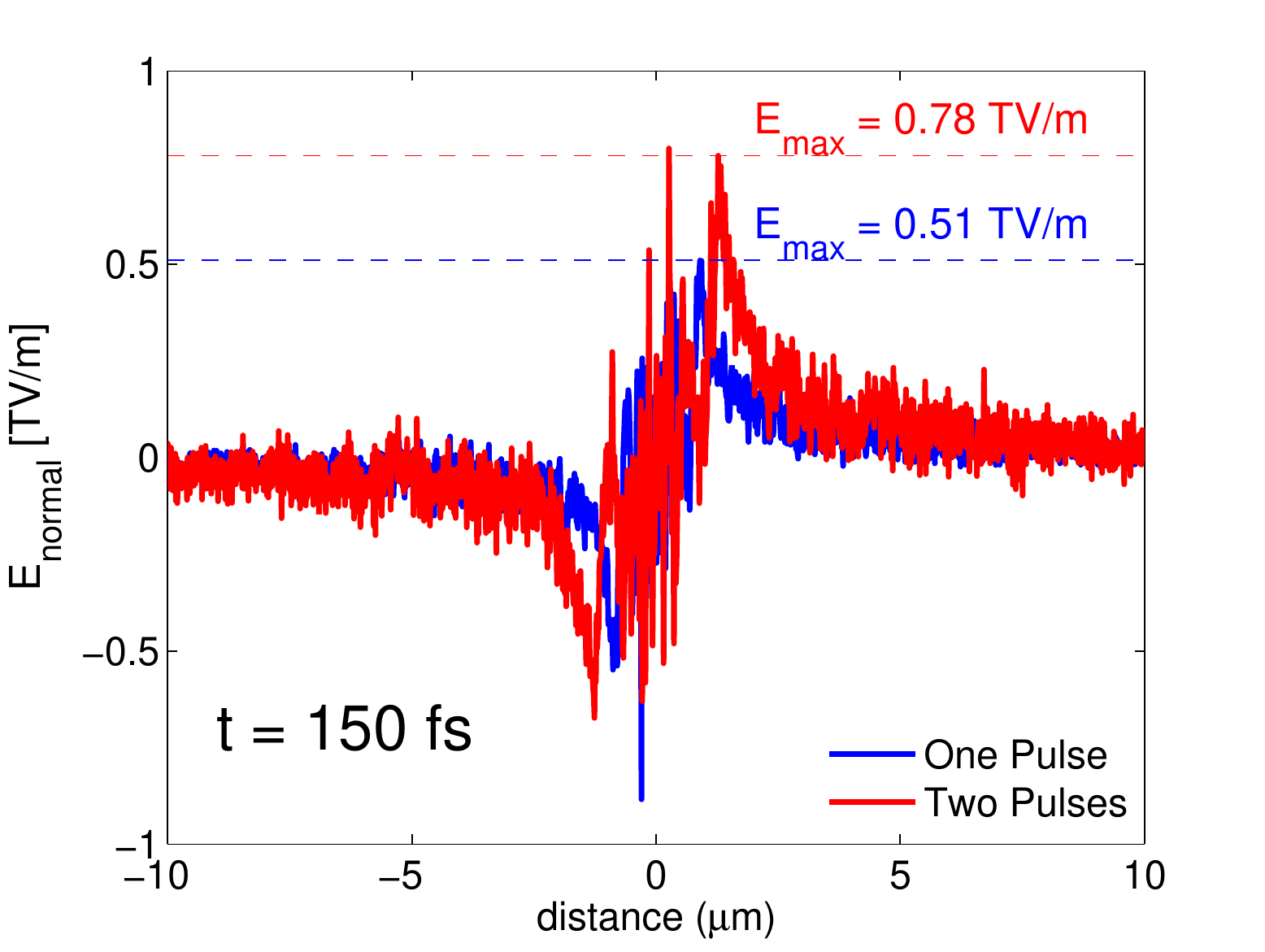}
\vspace{-0.5cm}
\caption{Upper panels show electron energy spectra for 100~fs (left panel) and 150~fs (right panel) into the simulation. Lower panels show a lineout of the electric field perpendicular to the target with the left panel showing 100~fs and the right panel showing 150~fs. In the top panels the exponential fit (a.k.a. hot temperature) is shown with dashed lines and the ratio of electrons above 200~keV in energy is shown. In the bottom panels the maximum electric field on the side of the target away from the laser is highlighted with dashed horizontal lines.}\label{fig:espec_Efield}
\end{figure*}

\begin{figure*}
\includegraphics[width=7in]{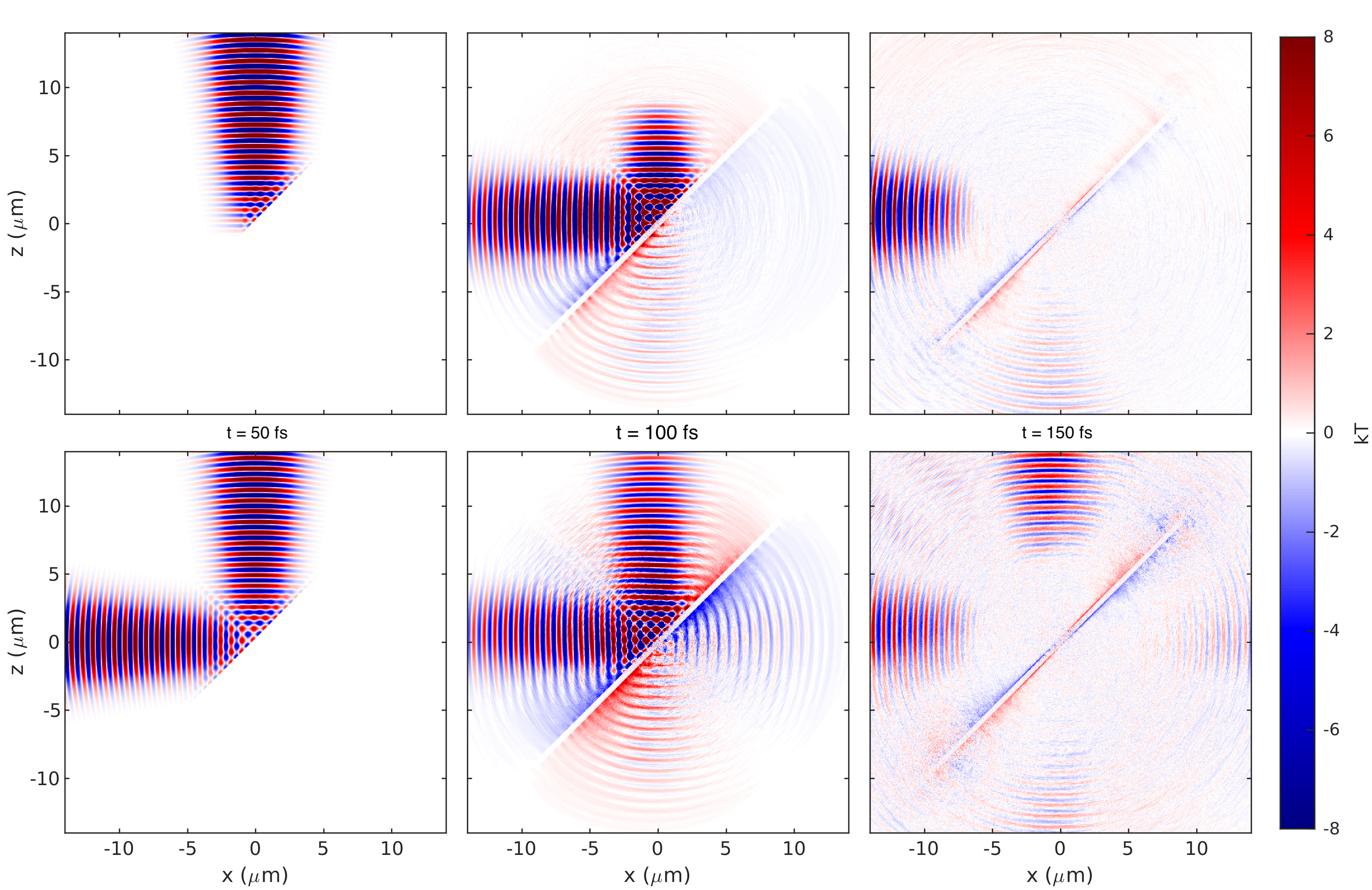}
\caption{Magnetic field component perpendicular to the plane of the simulation for the single pulse simulation (top panels) and double pulse simulation (bottom panels) at 50~fs (left panels), 100~fs (center panels), and 150~fs (right panels) }\label{fig:bfields}
\end{figure*}

\section{Results from Temporally and Spatially Well Aligned Pulses}
\label{sec:results}

Fig.~\ref{fig:efields} presents the electric fields normal to the target for the single pulse simulation (top panels) and the double pulse simulation (lower panels) at three different times. The first panel shows 50~fs after the beginning of the simulation when the pulses are beginning to interact with the target. The laser arriving from the top is polarized in the $x$ direction and moving in the $-z$ direction. The second laser pulse, which is polarized in the $z$ direction, is moving in the $+x$ direction towards the target.
Because the target normal E fields are plotted in Fig.~\ref{fig:efields} it is easy to see both pulses in the bottom panel and only one pulse in the top panel.

The middle panels of Fig.~\ref{fig:efields} show the electric field component normal to the target at 100~fs after the start of the simulation. At this time, the laser pulse is strongly interacting and reflecting from the target, producing constructive and destructive interference. Only a small fraction of the pulse energy is transmitted through the target. Interestingly, the sheath fields at the back side of the target are highly symmetric in the double pulse simulation whereas the single pulse simulation is less symmetric and features a weaker electric field.

The right panels show at a time 150 fs after the start of the simulation, at which time the laser pulses have passed the target and are leaving the simulation. The sheath field in the double pulse simulation remains larger than the single pulse simulation and more symmetric. 
At the center of both of these panels, the target has visibly expanded due to the TNSA process.

Figure~\ref{fig:espec_Efield} shows the electron energy spectra at 100~fs and 150~fs. At both times, electrons have a higher effective ``hot" temperature and are more abundant for the double pulse case. For example, there are roughly twice as many electrons above 200~keV energy in the the double pulse simulation. Figure~\ref{fig:espec_Efield} also includes lineouts of the electric field normal to the target showing noticeable enhancement of the accelerating field for the double pulse simulation by a factor of about 1.5.

Fig.~\ref{fig:bfields} shows snapshots of the magnetic field at 50, 100 and 150~fs. Specifically we show the component of the magnetic field perpendicular to the plane of the simulation ($B_y$).

\begin{figure*}
   \includegraphics[width=7in]{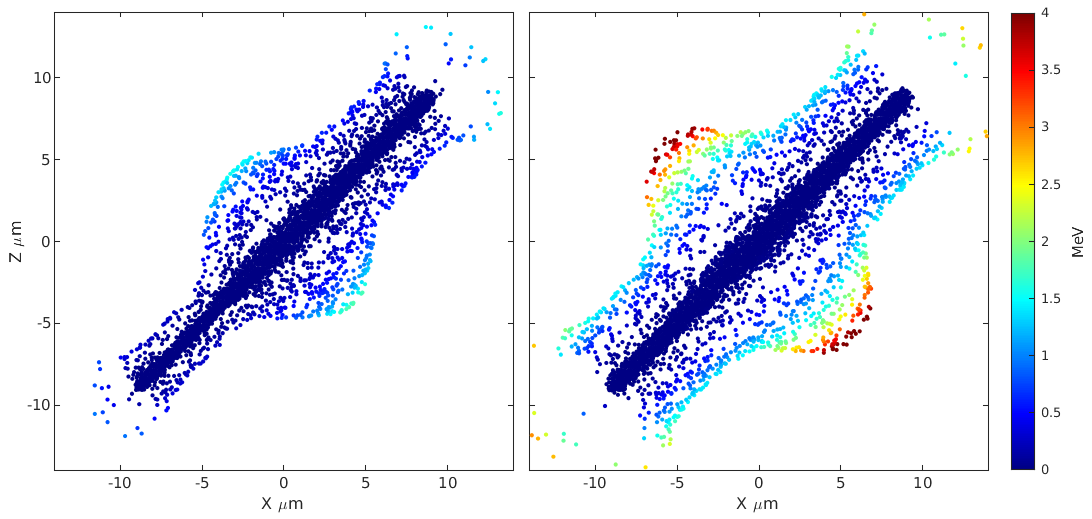}
   \vspace{-0.5cm}
   \caption{(left panel) Proton energy 450~fs after pulse for single pulse. (right panel) Proton energy 450~fs after pulse for double pulse.}
   \label{fig:escatter}
\end{figure*}

Fig.~\ref{fig:escatter} shows the positions and energies of a random selection of protons in our simulation 450~fs after the start of the simulation. The left panel is the single pulse simulation and the right panel is the double pulse simulation. We highlight 450~fs because the difference between the single pulse and double pulse simulations are particularly apparent at this time but it is still too early for protons to begin leaving the simulation. As the plot shows, the double pulse simulation has higher maximum proton energies than the single pulse simulation and these protons naturally travel further from the target by 450~fs.

\begin{figure*}
\includegraphics[width=2.5in]{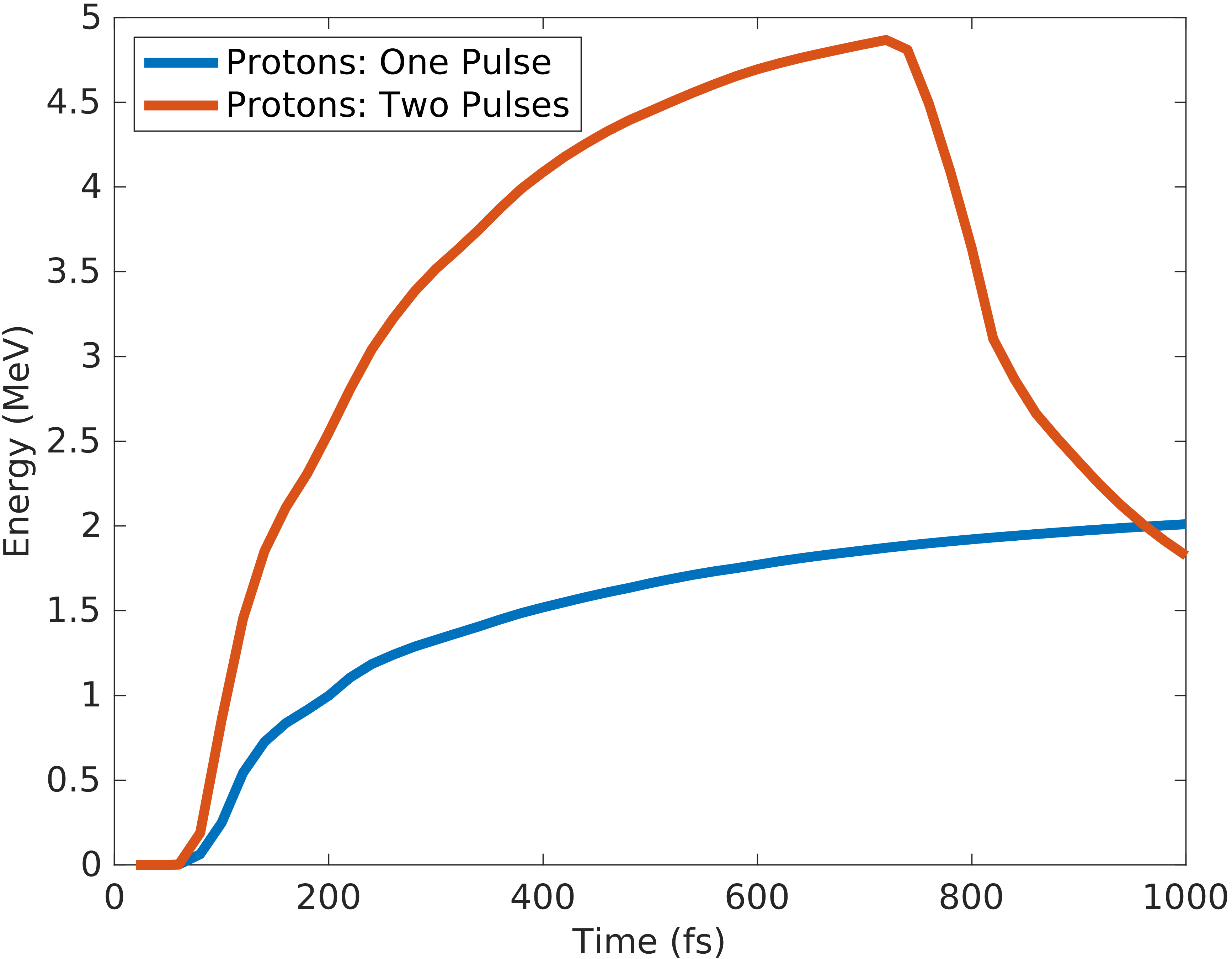}\includegraphics[width=2.7in]{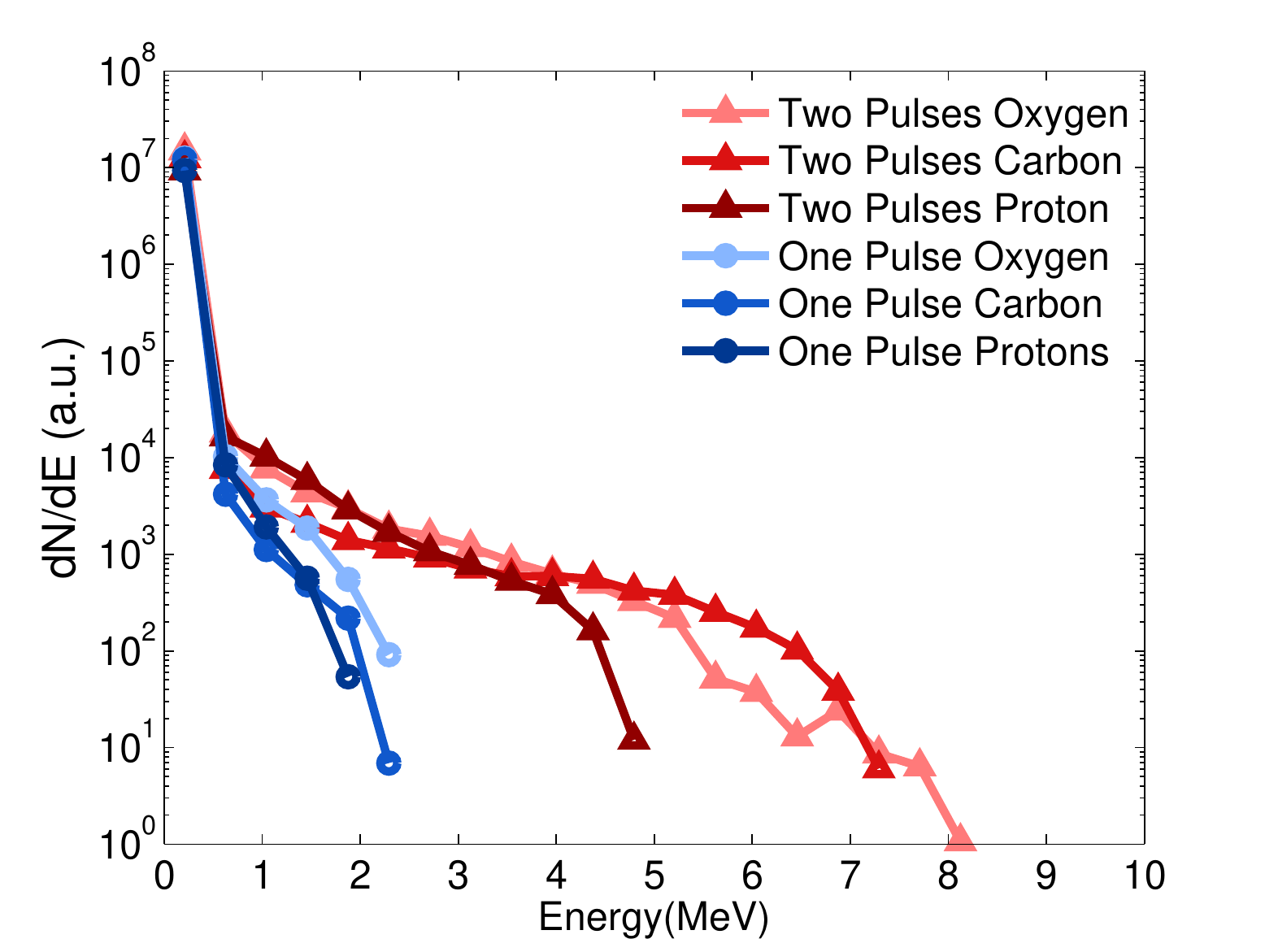}
\vspace{-0.25cm}
\caption{(left panel) Average energy of the 1000 most energetic proton macroparticles in the single pulse and double pulse simulations. (right panel) Comparing ion spectra at 700~fs after the beginning of the simulation.}
\label{fig:energy}
\end{figure*}

The maximum proton energy in the simulation changes with time so it is important to illustrate this. The left panel of Fig.~\ref{fig:energy} shows the average energy of the top 1000 proton macroparticles in the single and double pulse simulations. 
This figure shows a significantly sharper increase in the maximum proton energies of the double pulse simulation compared to the single pulse simulation. After 700~fs, the most energetic protons in the double pulse simulation begin to leave grid which is why the energies drop significantly. At 700~fs into the simulation, the average energy of the top 1,000 protons was 1.85~MeV for the single pulse simulation and 4.84~MeV for the double pulse simulation. This is a remarkable increase of 2.9 times greater proton energy, despite both simulations having the same total pulse energy. 

The right panel of Fig.~\ref{fig:energy} shows the proton and ion spectra for both simulations at 700~fs. These spectra show an increase in the maximum proton energy from $\sim$2~MeV to $\sim$5.5~MeV as well as increases in the numbers of energetic protons and ions. Note that because the left panel of Fig.~\ref{fig:energy} is the average of the 1000 most energetic proton macroparticles that these average energies there are slightly lower than the peak energies seen in the energy spectra in the right panel of Fig.~\ref{fig:energy}.

\begin{figure*}
\includegraphics[width=3.5in]{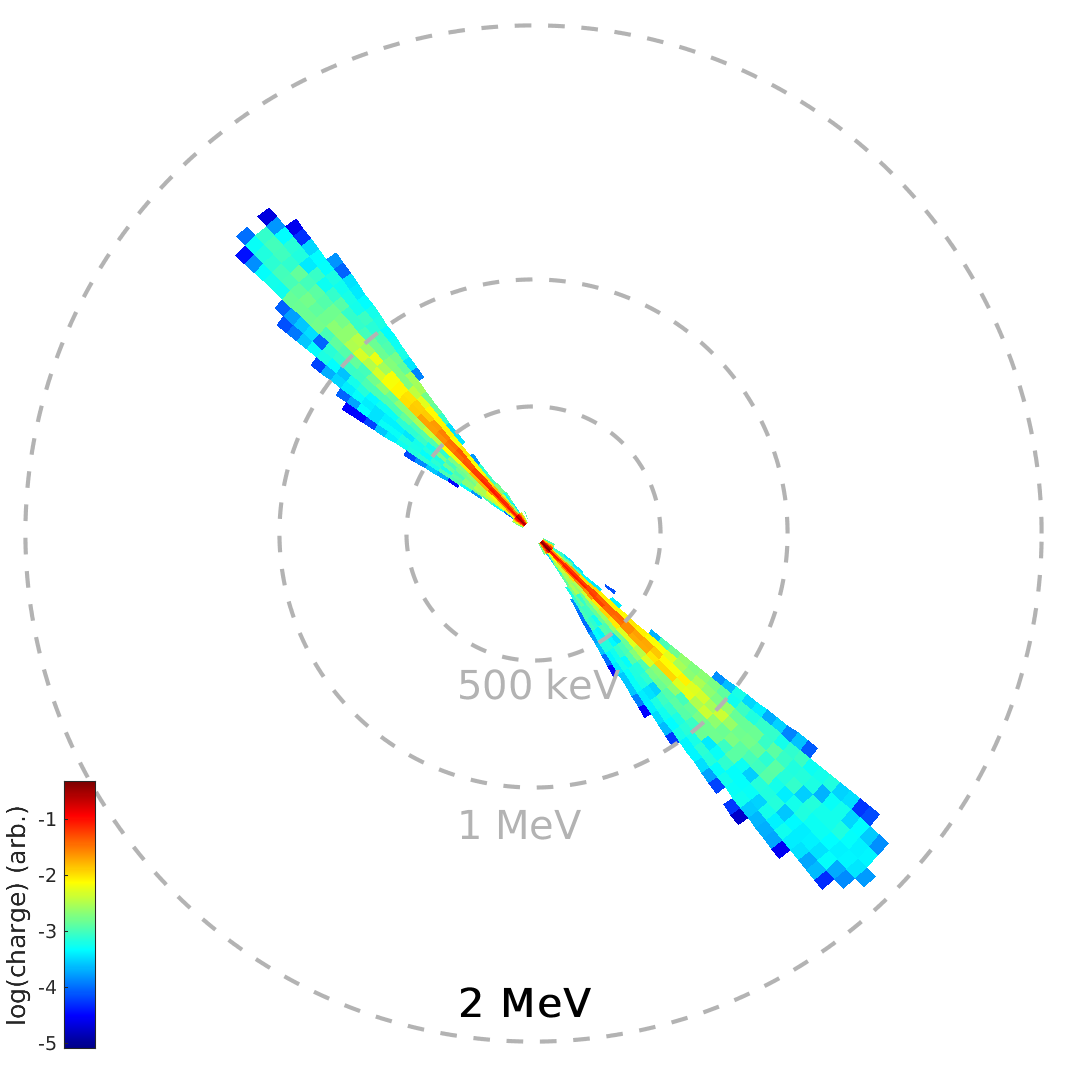}\includegraphics[width=3.5in]{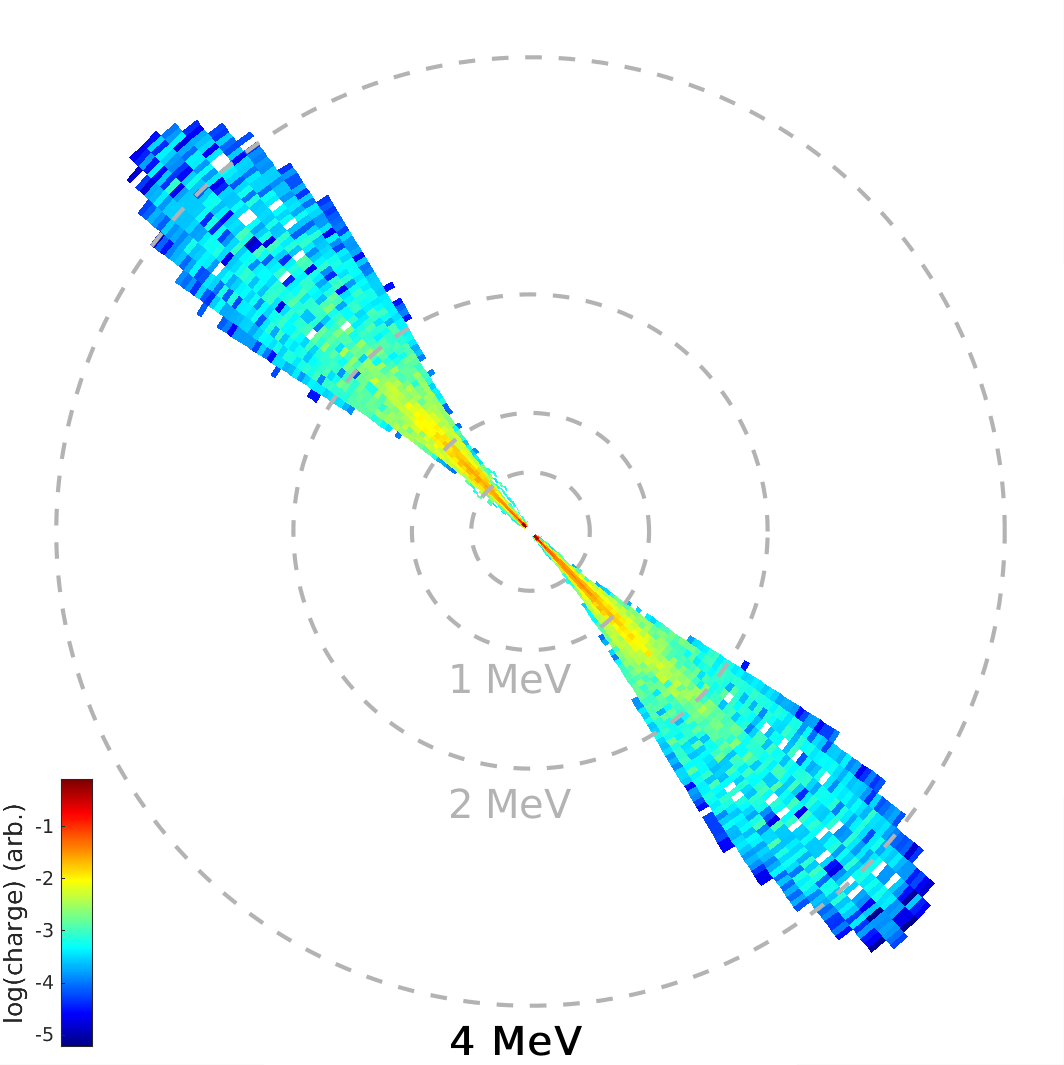}
\vspace{-0.25cm}
\caption{An analysis of the ejection angles and energies of protons in the single pulse (left panel) and double pulse (right panel) simulations at 450 fs. The protons are binned both by ejection angle and energy with the color proportional to the log of the number of proton macroparticles in that bin. Forward going protons appear in the bottom right of the plot while back-directed protons appear in the top left. Note that the scales for the energy are different across the two figures.
}
\label{fig:collimation}
\end{figure*}

Fig.~\ref{fig:collimation} compares the ejection angles and energies of the protons in both simulations at 450~fs calculated from the momenta of the proton macroparticles. Larger energy particles are further from the origin at the center of the plot. The TNSA process accelerates protons both in the forward direction from the target as well as the back (specular) direction. In Fig.~\ref{fig:collimation}, forward moving protons appear in the bottom right while back directed protons appear in the top left. 

At 450~fs into the simulation, 79\% of the protons above 300~keV in the single pulse simulation were found to have an ejection angle within $\pm5\degree$ from normal to the target, compared to 69\% for the double pulse simulation. Considering only the 1,000 proton macroparticles with the greatest energies at 450~fs, the percent of proton macroparticles which were found to have an ejection angle between within $\pm5\degree$ from normal to the target was 86\% in the single pulse simulation and 72\% of the double pulse simulation. These 1000 most energetic proton macroparticles had an energy range between 1.41~MeV to 1.91~MeV in the single pulse simulation and from 3.83~MeV to 4.87~MeV in the double pulse simulation. Although the ions in the single pulse simulation are somewhat more collimated than the double pulse sim, it is important to note that there are significantly more protons with energies greater than 300~keV in the double pulse simulation as discussed in the next section.

\begin{figure*}
\includegraphics[width=3in]{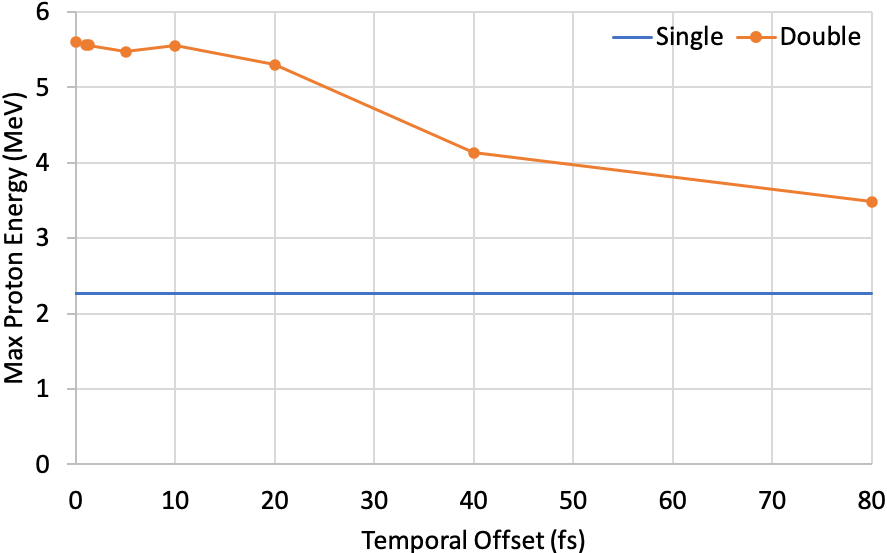}\includegraphics[width=3in]{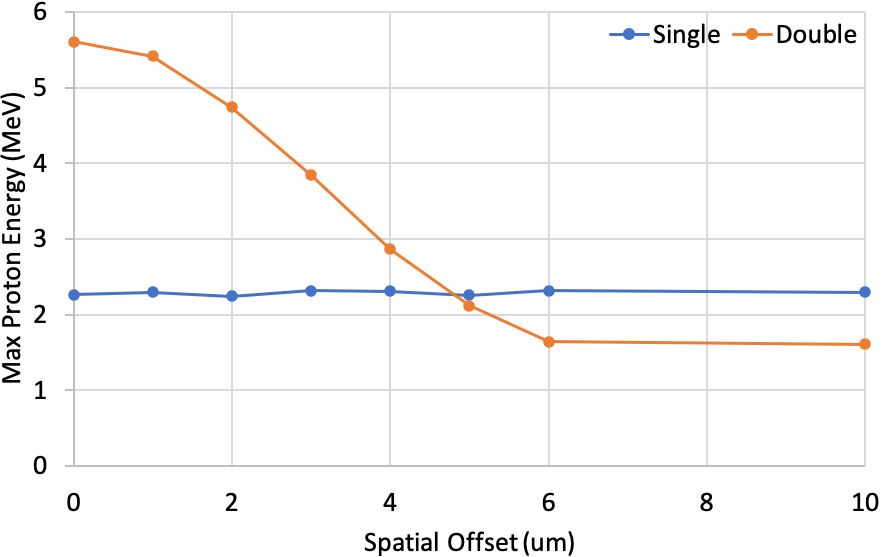} \\
\includegraphics[width=3in]{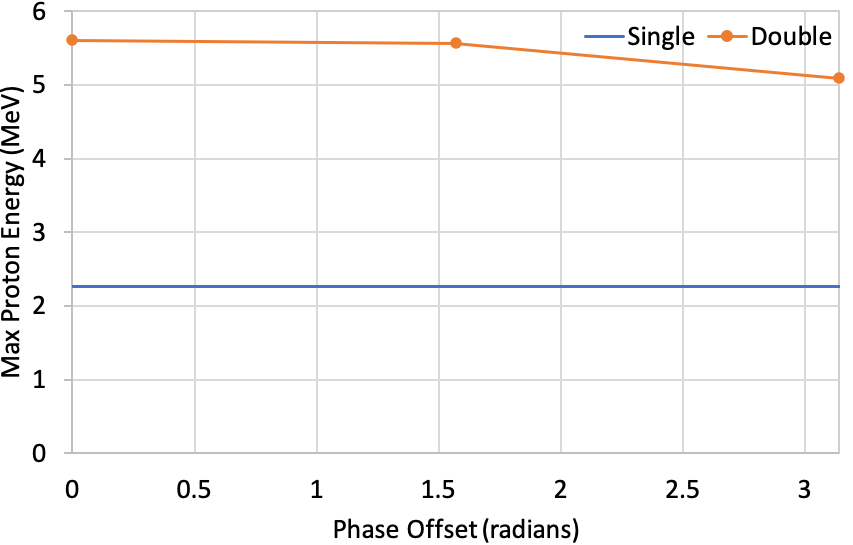}
\vspace{-0.25cm}
\caption{Maximum proton energy at 750~fs after the start of the simulation for various temporal offsets (top left), spatial offsets (top right), and phase offsets (bottom). The top left panel shows temporal offsets ranging from 1~fs to 80~fs. The double pulse simulations had one pulse delayed and the other pulse with no delay. The maximum proton energy for the single pulse simulation is shown with a horizontal line. The top right panel shows spatial offsets ranging from 1~\si{\micro\metre} to 10~\si{\micro\metre}. 
The bottom panel shows phase offsets ranging from 0 to $\pi$ radians. 
}
\label{fig:OffsetEnergy}
\end{figure*}

\section{Results from Temporally or Spatially Misaligned Pulses}
\label{sec:scan}

An important question for experimental demonstration is how closely the two beams must overlap, both spatially and temporally, to see a significant increase in peak ion energy and whether the phases of the two beams matter. In experiment, aligning two beams with high precision can be challenging and the beams may shift slightly between shots. These shot-to-shot variations (sometimes called jitter) can cause inconsistently aligned pulses. With this in mind we performed a number of additional 2D3v PIC simulations using the same setup described in Sec~\ref{sec:setup} but with spatial, temporal or phase offsets. 

Fig.~\ref{fig:OffsetEnergy} shows the maximum proton energy from additional 2D3v PIC simulations using the same setup described in Sec~\ref{sec:setup} but with spatial, temporal or phase offsets. The top left panel of Fig.~\ref{fig:OffsetEnergy} shows the effect of temporal offsets and the top right panel of Fig.~\ref{fig:OffsetEnergy} shows the effect of spatial offsets on the maximum proton energies at 750~fs after the start of the simulation. By temporal offset we mean that one of the beams had a time delay. By spatial offsets we mean that both beams were moved evenly away from the center of the target without changing the angle of incidence. Both the single and double pulse beam positions were further adjusted to ensure that the peak focus (i.e. maximum intensity) remained at the target despite this change. The distance between the peak focus of the two beams is represented in the figure (i.e. the distance from the center of the target to the peak focus is half the distance represented in the figure). Both upper panels show a decrease in peak ion energy as the two beams are further from perfect temporal and spatial alignment. We also performed two additional simulations with perfect temporal and spatial overlap where the phase was offset in one of the beams. 

Fig.~\ref{fig:OffsetEnergy} shows that the double pulse scheme still produces approximately a factor of two enhancement in max proton energy even for pulses that are 40~fs delayed or that are spatially offset by about 3~$\mu$m. We did not perform simulations with both a spatial and temporal offset even though real experiments will likely have some degree of misalignment in both respects. The two simulations we performed to examine phase offsets found that the maximum proton energy only decreased (at most) by 10\%.

\section{Analysis and Discussion}
\label{sec:discussion}

The stated aim of this paper is to simulate a potential demonstration experiment of ``enhanced TNSA" which is a phenomenon described by \citet{Ferri2019} and \citet{Ferri_etal2020}. In this section we comment on the nature of the enhancement as seen in our simulations which we include as an important background for what a potential demonstration experiment should try to measure. We also summarize some of the comments on the nature of the enhancement mentioned in  \cite{Ferri2019,Ferri_etal2020}.  Additional considerations for the required spatial and temporal alignment of the two pulses and robustness to phase differences are also discussed.

\begin{figure*}
\includegraphics[width=2.9in]{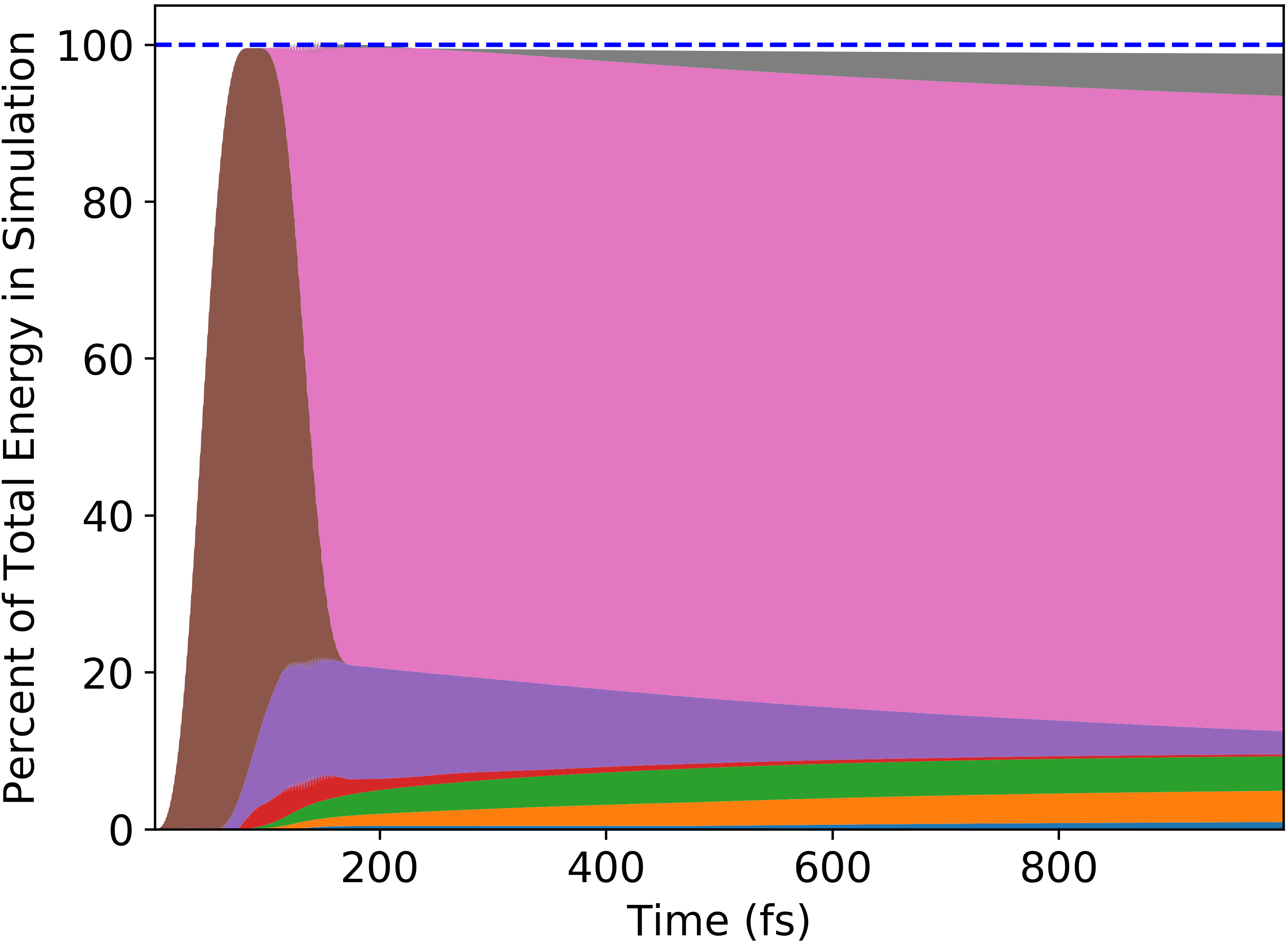}\includegraphics[width=2.9in]{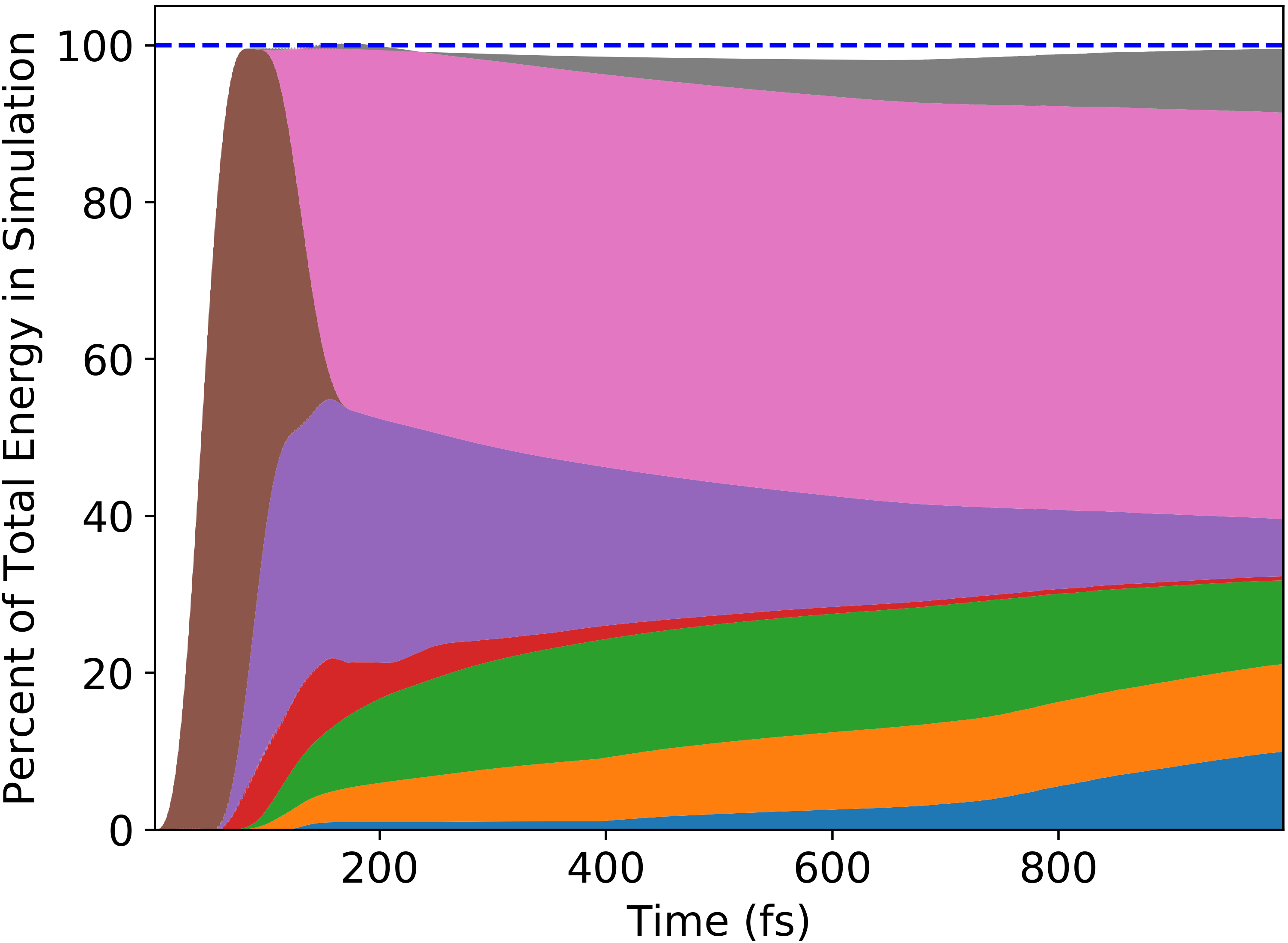}\includegraphics[width=1.2in, trim=0 0 0 -1in]{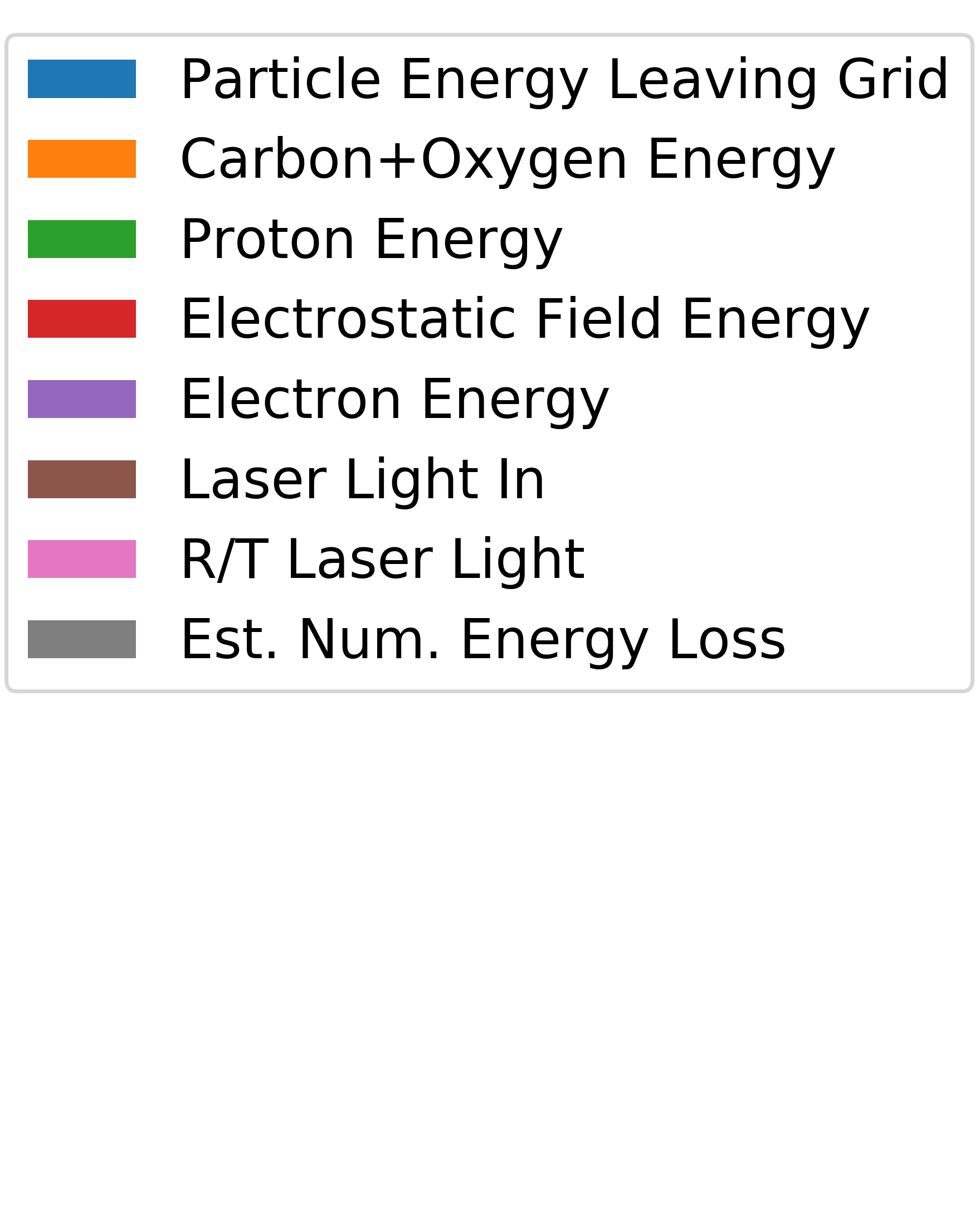}
\vspace{-0.25cm}
\caption{Evolution of various sources of energy in the Single Pulse (left panel) and Double Pulse (right panel) simulations. A horizontal dashed line shows the analytically derived total laser energy that enters the simulation. The LSP code includes an approximate estimate of energy loss due to numerical effects which is shown on the plot. Also included is an estimate of the laser energy that leaves the simulation through reflection or transmission. The lines do not all add up to 100\% likely due to the approximate nature of the estimate for energy loss due to numerical effects in the LSP code.} 
\label{fig:StackedEnergy}
\end{figure*}

We compared the evolution of the energies in the two simulations from Sec.~\ref{sec:results} in Fig.~\ref{fig:StackedEnergy}. We created this plot from the reported total energy in particles of various species versus time in the simulation, from measurements of the total ``field energy" which is the sum of energy in electric fields and magnetic fields everywhere in the simulation versus time and from measurements of the electromagnetic energy leaving the grid. Our simulations also include an approximate estimate for energy loss due to purely numerical effects like the damping of the electromagnetic fields from the implicitness used by the LSP code. The sum of all these energy components add up to nearly the total energy of the laser energy entering the simulation (horizontal dashed line). Deviation from 100\% is likely due to the approximate nature of the model that determines the estimated energy loss due to numerical effects. The estimated numerical energy loss were similar with reported 5.43\% energy loss for the single pulse simulation and 8.15\% for the double pulse simulation after 1~picosecond of simulated time. 

The difference between the left and right panels of Fig.~\ref{fig:StackedEnergy} is striking. According to these data, 21\% of the laser light is absorbed in the single pulse simulation while 53\% of the light is absorbed in the double pulse case. This increased absorption ultimately produces 2.25 times more energy in electrons. So we conclude that a demonstration experiment should see a significant increase in absorption and corresponding decrease in reflected laser light.

Fig.~\ref{fig:espec_Efield}, which shows the electron energy distributions in the upper panels, is insightful for understanding the difference in total electron energy in the two simulations. Both the numbers of energetic electrons and the energies of electrons are enhanced. According to \citet{Ferri2019} the reason for the enhanced electron acceleration (which of course is related the increased absorption) is due to better constructive interference of the laser electric fields on the target. 
As they illustrate, in the double pulse case the two beams constructively interfere even before reaching the target while the single pulse case can only interfere with its reflection (which is attenuated due to the absorption). Ultimately the strongest constructive interference occurs for the double pulse case. According to Eq.~1 in \cite{Ferri2019}, the constructive interference should enhance the electric field by a factor of $\sqrt{2}$, ultimately doubling the effective laser intensity near the target. We can examine whether the measured hot temperatures ($T_{\rm hot}$) from our simulations are consistent with this result. Using ponderomotive arguments for $T_{\rm hot}$ (e.g. Eq.~6 from \cite{Macchi_etal2013}) with $a_0 = 1.5$ for the single pulse simulation, the expected hot temperature for electrons is 0.23~MeV. Twice this intensity would imply $a_0 = 2.1$ and the same formula would predict a hot temperature of 0.41~MeV. These numbers are very similar to the values we obtain in the upper panels of Fig.~\ref{fig:espec_Efield}. Other scaling formulae (cf. \cite{Haines_etal2009} and references therein) give qualitatively similar results.

Another interesting consistency check that can be done with Fig.~\ref{fig:espec_Efield} is to see if the enhanced electric field that develops on the target is consistent with expectations from TNSA theory. As is well known (e.g. \cite{Mora2003,CERN}), according to 1D arguments, the maximum electric field $E_{\rm max}$ should scale as $k_B T_{\rm hot} / e \lambda_D$ where $\lambda_D$ is the Debeye length and the Debeye length should scale as $\sqrt{k_B T_{\rm hot} / n_{\rm hot}}$  where $n_{\rm hot}$ is the density of hot electrons. This predicts a scaling of $E_{\rm max} \sim \sqrt{n_{\rm hot} k_B T_{\rm hot}}$. Fig.~\ref{fig:espec_Efield} includes a measurement of the ratio of the hot electron density for electrons above an energy threshold of 200~keV. (The energy threshold of 200~keV is similar to the ponderomotive energy.) Broadly we find that hot electron density and the hot electron temperature are both enhanced by a factor of two which predicts an increased $E_{\rm max}$ by a factor of $\approx \sqrt{2 \cdot 2} = 2$. The enhancement in $E_{\rm max}$ from the bottom panels of Fig.~\ref{fig:espec_Efield} is measured to be close to 1.5. We note that using a 50\% lower electron energy threshold reduces the enhancement in $n_{\rm hot}$ without significantly changing $T_{\rm hot}$, bringing the predicted and measured enhancement in $E_{\rm max}$ into reasonable agreement. We regard these results as qualitative evidence that TNSA is the dominant ion acceleration mechanism in both simulations. 

There is an interesting discussion in \citet{Ferri_etal2020} that proposes that differences in the structure of the quasi-static magnetic fields that develop on the target is another important reason why ion acceleration is more efficient for the double pulse case. In future work we intend to consider this question in depth with particle tracking. For the present work, it bears mentioning that the magnetic fields that develop on the target in our simulations (Fig.~\ref{fig:bfields}) have a structure that greatly resembles magnetic fields shown in \citet{Ferri_etal2020} which were obtained from higher intensity simulations. This raises our confidence that double pulse experiments at $\sim 5 \cdot 10^{18}$\si{W.cm^{-2}} intensities on milliJoule class laser systems can provide useful insights to anticipate how experiments at much higher intensities would perform. 

In terms of the still-unverified hypothesis that quasi-static magnetic fields play an important role in producing the enhancement we offer one insight: our double pulse simulations are more effective at converting electron energy to ion energy than the single pulse simulations. Because the ion acceleration happens on a similar or longer timescale that quasi-static magnetic fields develop it is plausible that magnetic fields are key to this result. In the single pulse simulations we found that 28\% of the electron energy is ultimately converted to proton energy with another 26\% going to Oxygen and Carbon. In the double pulse simulations 43\% of the electron energy is  converted to proton energy and 32\% of the electron energy is converted to Oxygen and Carbon ions. So not only is there more energy in electrons in the double pulse simulations (as mentioned earlier), a higher fraction of this electron energy serves to accelerate ions. Ultimately the 2.25x increased energy in electrons combined with the enhanced transfer of that energy to ions produces a factor of 3.4 times as much energy in protons and a factor of 2.8 times as much energy in Oxygen and Carbon in the double pulse simulations. As mentioned earlier, the maximum energies of the ions are also significantly increased by a factor of 2.5-3 (Fig.~\ref{fig:energy}). Interestingly, this is an even larger enhancement than was seen in \cite{Ferri2019,Ferri_etal2020}. A possible explanation for this could be that \citet{Ferri2019,Ferri_etal2020} both simulate already fully ionized targets whereas our simulations include a prescription for ionizing ion species through field ionization as discussed in Sec.~\ref{sec:setup}. We performed simulations with fixed ionization (not shown), keeping all species singly ionized, and found that the enhancement in the numbers of protons and peak proton energy in the double pulse simulation compared to the single pulse simulation was less than what we found in Figs.~\ref{fig:energy} for the ionization simulations and closer to the factor of two that \citet{Ferri2019,Ferri_etal2020} found.

A reasonable question is whether our results for the enhancement are biased by the finite width of our target. We performed two additional simulations (one single pulse, one double pulse) with a simulation grid that was twice the size (56~$\mu$m $\times$ 56~$\mu$m) as our fiducial simulations, with twice the width of the target (51~$\mu$m) and with twice the cells in both directions (5600 $\times$ 5600). Because of the additional computational expense, these simulations were only run to 650~fs. Besides these details, the the simulation setup was identical to the simulations described in Sec.~\ref{sec:results}. Broadly, these simulations yielded qualitatively similar results to the fiducial simulations described in Sec.~\ref{sec:results}. At 650~fs, in the single pulse simulation with the larger grid the average energy of the 1000 most energetic proton macroparticles was 1.50~MeV. For the larger double pulse simulation the same measurement yields 4.14~MeV. For comparison, the simulations with the smaller target and grid yield 1.77~MeV (single) and 4.71~MeV (double) for the same amount of time after the front of the laser pulse reaches the target. The larger grid therefore reduces the maximum proton energies by between 14 and 18\% but we still find that the ratio of the maximum proton energies in the simulations is relatively unchanged. This leads us to conclude that our qualitative conclusions would be unchanged if our simulations had used the larger grid and target throughout this paper.

In an effort to provide useful information to experimentalists regarding the possibility that the two pulses may be temporally or spatially misaligned we performed a number of additional simulations as presented in Sec.~\ref{sec:scan} and highlighted in Fig.~\ref{fig:OffsetEnergy}. Those simulations found that ion acceleration remained enhanced (compared to a single pulse simulation) by roughly factor of two even for temporal offsets of 40~fs. Given that the nature of the effect requires the constructive interference of the two pulses it is unsurprising that the effect is diminished when the delay becomes comparable to the 42~fs FWHM duration of the pulses. At 80~fs, a smaller enhancement still remains. In Sec.~\ref{sec:intro} we mentioned studies where enhanced ion acceleration is found from two pulses arriving on target from the same angle but with a time delay\cite{Markey_etal2010,Brenner_etal2014,Ferri_etal2018,kumar_gupta_2020}. Enhancement can potentially occur if the first pulse creates a preplasma on the target that increases the coupling of energy from the second pulse. A similar physical mechanism is explored, at higher intensities, in \citet{Kim_etal2020}.

Another set of simulations presented in Sec.~\ref{sec:scan} found that pulses that were temporally aligned (zero delay) but with a spatial offset between 2~$\mu$m and 3~$\mu$m still enhanced the ion energies by about a factor of two compared to a single pulse simulation. In our simulations the spot sizes were both 2.2~$\mu$m FWHM. Our results confirm intuition that aligning the laser spots within the FWHM of the spot size is important for producing the enhancement.

For brevity we did not perform simulations with both a spatial and temporal offset even though spatial and temporal offsets are likely on some level in any real demonstration experiment. Sec.~\ref{sec:scan} does present simulations with different phase offsets between the two pulses, finding that, for our parameters, the maximum proton energy is robust to phase differences.

Other important questions remain including to what degree the enhancement is robust to the presence of a pre-plasma. We explore this some in Appendix~\ref{ap:preplasma} but more attention is needed, especially because there are potential implications for the required pre-pulse contrast in the experiment. 

Another concern is to what extent the 2D3v geometry of our simulations affect our results. 
\citet{Ferri2019} perform two 3D simulations and found reasonable agreement with their 2D3v results. It is unclear if this will be true in general.

There are also a number of questions about the role of the quasi-static magnetic fields that particle tracking would illuminate. We have begun this analysis (not shown) but find that the substantial difference in absorption in the two simulations makes it difficult to pinpoint the role of the magnetic field separately from the role of the increased absorption. Another interesting investigation would be to consider three laser pulses (for the same total energy) as proposed recently by \citet{Jiang_etal2020}. We defer these questions for future work.

\section{Conclusion}
\label{sec:conclusion}

Using 2D3v PIC simulations we consider if a scenario for double pulse enhanced TNSA, as described by \citet{Ferri2019} and \citet{Ferri_etal2020}, would be feasible on a milliJoule class laser system. For a number of practical reasons, at the time of this writing their idea of enhancing ion acceleration using two laser pulses with half energy (and intensity) and a wide ($\sim 45^o$ angle of incidence) has not yet been convincingly demonstrated experimentally even though simulations predict substantially enhanced ion acceleration. 
We performed simulations with laser pulses that were 3.5 times less intense than the lowest intensity investigated in \citet{Ferri2019} and we used a different target material and target thickness than was used there. We assumed laser parameters similar to the milliJoule class laser system described in \citet{morrison_etal2018} which previously demonstrated up to 2.5 MeV protons.
Comparing two pulses with intensities of $2.5\cdot 10^{18}$~\si{W.cm^{-2}} on target to a single $5\cdot 10^{18}$~\si{W.cm^{-2}} pulse with the same pulse duration and spot size, our simulations find enhanced peak proton energy and conversion efficiency from laser energy to proton and ion energy by about a factor of 3 and with a similar divergence for protons. Interestingly, this factor of 3 enhancement is larger than the overall factor of 2 enhancement described in \citet{Ferri2019}. 

We also find that the results for the electric and magnetic fields in our simulations qualitatively resemble snapshots of the electric and magnetic fields presented in \citet{Ferri2019} and \citet{Ferri_etal2020} from simulations performed at higher intensities. This implies that the constructive interference and quasi-static magnetic field generation occurs analogously at our lower intensity. We also present results from a series of simulations that examine the spatial or temporal misalignment of the two laser pulses or phase differences.

Investigation into the nature of the enhancement seems to confirm that TNSA is the dominant acceleration mechanism and that constructive interference provides a reasonable explanation for the enhancement in the electron energy distribution.

Our results lead us to the conclusion that milliJoule class laser systems can play an important role in demonstrating double pulse enhanced TNSA. Such an experiment should be insightful for later experiments at much higher intensities and energies.

\begin{acknowledgments}
This research was supported in part by an appointment to the Postgraduate Research Participation Program at the U.S. Air Force Institute of Technology, administered by the Oak Ridge Institute for Science and Education through an interagency agreement between the U.S. Department of Energy and AFIT. This research is also supported by the Air Force Office of Scientific Research under LRIR Project 17RQCOR504 under the management of Dr.~Andrew Stickrath.  NR acknowledges summer support from the OSU SCARLET laser group. Simulations were performed on the ASC Unity cluster at Ohio State University and the Ohio Supercomputer Center~\cite{OhioSupercomputerCenter1987}.

\end{acknowledgments}

\section*{Data Availability}
The data that support the findings of this study are available from the corresponding author upon reasonable request.

\onecolumngrid
\appendix

\section{Effects of a Preplasma}
\label{ap:preplasma}

Ultra-intense laser experiments typically involve some level of unintended laser light reaching the target in the picoseconds to nanoseconds before the arrival of the main ultra-intense pulse, which can produce a plume of material in front of the target referred to as a ``preplasma". An interesting question with experimental implications is how robust the enhancement is to the presence of a preplasma. To explore this, we performed simulations with the same grid and laser parameters as described in Sec. \ref{sec:setup}, changing only the target. The target was modified to have a preplasma with an exponential density profile with a scale length of 0.4~\si{\micro\metre}. This scale length was chosen to be similar to the optimal scale length for maximizing the proton energy found in \citet{Ferri2019}. The target profile was extended until the density of electrons reached $10^{20}$~cm$^{-3}$ with other species in proportion to ensure the target remained electrically neutral. This preplasma was added only to the face of the target being directly irradiated by the laser. The target was shifted such that the laser focus was at the critical density of the target, which initially is singly ionized.

\begin{figure*}[h]
\includegraphics[width=3.5in]{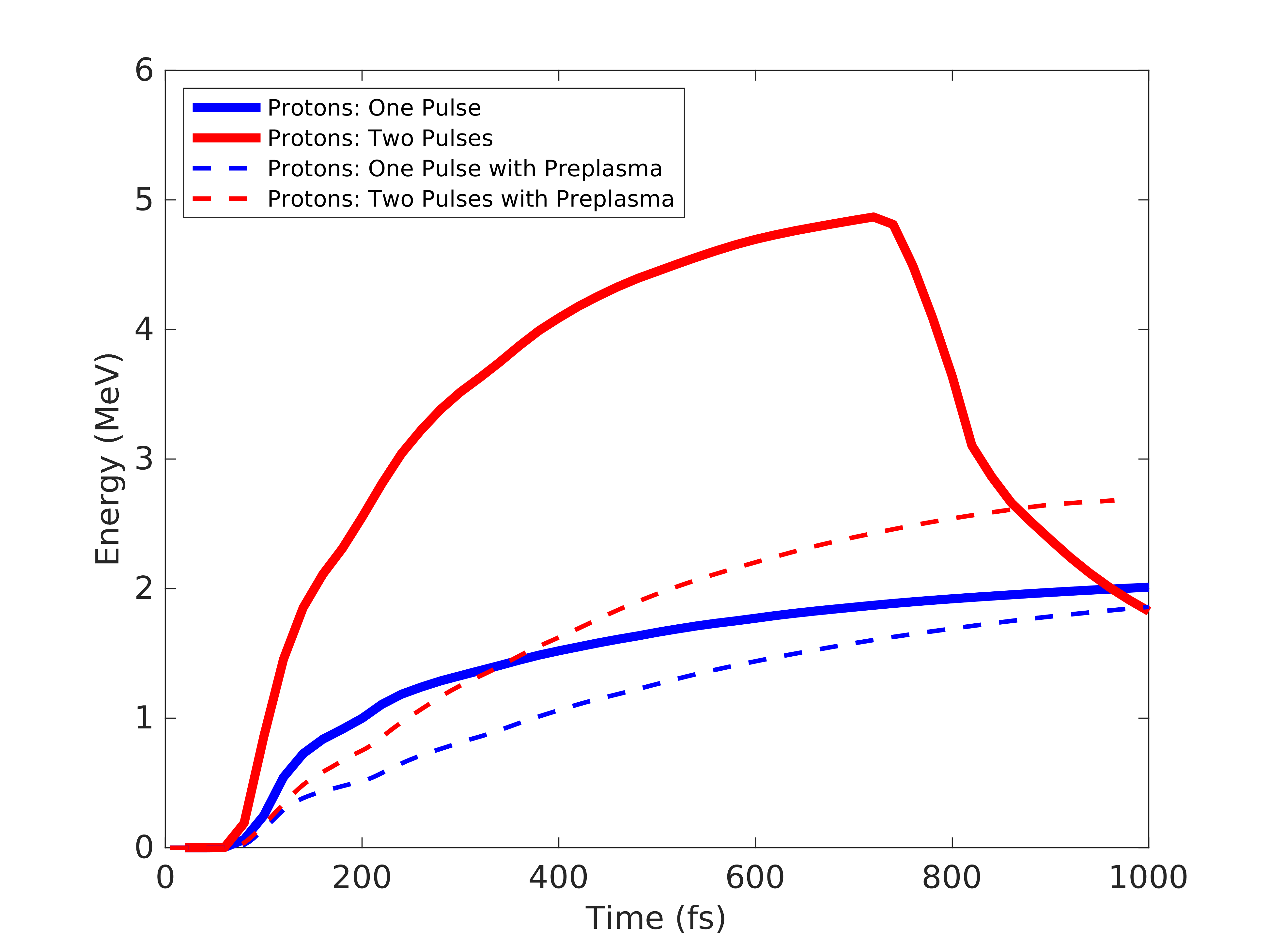}
\vspace{-0.25cm}
\caption{Average energy of the 1000 most energetic proton macroparticles in the single pulse and double pulse simulations with and without a preplasma.} 
\label{fig:PreplasmaEnergy}
\end{figure*}

Fig. \ref{fig:PreplasmaEnergy} shows the evolution of the average energy of the 1,000 most energetic proton macroparticles in each simulation. The presence of a preplasma only slightly reduces the proton energies in the single pulse results. The double pulse simulation still produces higher proton energies than the single pulse but by a much smaller factor -- instead of an enhancement of about 2.5, the enhancement with a preplasma is closer to 1.5. Due to the lower energies, the protons move more slowly and even the most energetic protons do not leave the simulation box by the end of the simulation, which is why we no longer see a sharp drop in the maximum energies around 700~fs.

This significant reduction in the maximum energies in the double pulse simulation is puzzling given that \citet{Ferri2019} found that the presence of a preplasma \emph{increased} the maximum proton energy. Given the many differences between the target, simulation and laser parameters investigated by \citet{Ferri2019} and our study, at this stage it is difficult to pinpoint exactly why we see the opposite result. The slight decrease in the peak proton energy from the single pulse result may be due to the effectively thicker target that the laser is interacting with due to the preplasma, and known trends for thicker targets reduce the peak proton energy (e.g. \cite{Kaluza_etal2004}). The reduction of the peak proton energy for the double pulse result likely relates to the way that the preplasma changes the interference pattern, making it less ideal than, for example, the analytic result for the electric and magnetic fields from reflection of a plane wave from a sharp interface described in Eq.~1 in \citet{Ferri2019}. 

Delving more into the nature of the diminished proton energies in the simulations is beyond the scope of this paper but understanding the effect of a preplasma on double pulse enhanced TNSA should be an important priority for future theoretical work. Experimentalists may find an enhancement only for relatively low prepulse levels and steep preplasma density profiles, as our simulations suggest.






\bibliography{aipsamp}

\end{document}